\documentclass{sf2a-conf2010}
\usepackage{graphicx}
\usepackage{hyperref}
\usepackage[]{natbib}

\begin{document}
\TitreGlobal{SF2A 2010}
%
\title{A CHANDRA Census of c2d YSOs: Evolution of X-ray Emission}
\author{M. Hamidouche}\address{USRA - NASA Ames Research Center, MS N211-3, Moffett Field, CA 94035}
\author{M. Jacobson}\address{Astronomy Department, U. of Illinois at Urbana-Champaign, Urbana, IL 61801}
\author{L. W. Looney$^2$}

\runningtitle{YSO-Xrays}
%
\setcounter{page}{237}
\index{Hamidouche, M.}
\index{Young, E.}
\index{Marcum, P.}
\index{Krabbe, A.}

\maketitle
\begin{abstract}
We present an analytical study of a large sample of $\sim$109 young stellar
objects in the X-ray.  Our objects were detected in X-ray independent of age. Unexpectedly, the X-ray energy is somewhat correlated with the ages. It decreases with time and with column density, while it should increase. We conclude that the youngest protostars, Class 0/I, emit X-rays in the ~1-8 keV band. The deeply embedded
sources with the strongest accretion activity are detected in the hard-band
($>$ 2keV) only. Due to extinction, their soft X-rays are not detected. To
explain the decline in energy, we suggest that within a timescale of few
Myrs the corona cools down via the accretion material.

%
%
\end{abstract}
\begin{keywords}
Infrared: Stars - Stars: Magnetic Field, Pre-Main-Sequence, Protostars -  X-rays: Stars 
\end{keywords}
\section{Introduction}


X-ray emission of pre-main-sequence T Tauri stars has been extensively studied over the last decades. The observed X-rays are explained to be from the coronal emission \citet{preibish07}. In this study, we further investigate the X-ray emission processes of such objects and its evolution since their youngest ages. We present a study of a large sample of young stellar objects (YSO) in the infrared and X-ray to probe their X-ray energy regime evolution from the youngest embedded and strongly accreting Class 0 objects into the non-accreting Class III objects. We track their X-ray activity during their lifetime, and thus the role of the accretion.

\section{Analysis}

We cross-correlated the data archive of YSO from the
infrared Spitzer c2d legacy project \citep{evans03} with CHANDRA X-ray
observations from the web based archive \citep{ws07} ANCHOR (AN
archive of Chandra Observations of Regions of star formation). We spatially
matched 109 detected YSOs within 4'' from three star-forming regions : NGC 1333,
Serpens, and $\rho$-Ophiuchus. We determine their ages from infrared spectral indices $\alpha$ and classify them from Class 0 to III, using the color-color diagram with the four IRAC bands [3.6]-[4.5] and [5.8]-[8.0]. We used the SED online fitting tool developed by \citet{rob07} to deduce the envelope accretion rate and the age. We obtain the column density $n_e$ and the X-ray energy from ANCHOR X-ray data. 

\section{Discussion}

The column density decreases with the derived infrared index, or the more evolved Class II/III objects are less embedded. These results show that our study is robust since our parameters are consistent with their physical implication. Overall, the infrared index $\alpha$ is slightly correlated with the stellar age. On the other hand, it is well correlated with the accretion rate. Nevertheless, we should point out that we have much less data for the older YSO than for the youngest ones. They are also scattered in the plot and it is difficult to determine any correlation.

\section{Conclusion}

We deduce in this study of 109 YSO that they all emit X-rays independently of their ages. Unexpectedly, the X-ray energy is somewhat correlated with the infrared indices (Figure 1), and thus it is anti-correlated with the ages. It decreases with time and column density, while it was expected to increase. We find that the youngest YSO protostars (Class I) emit X-rays in the $\sim$1-8 keV band. Interestingly, we do not see an apparent evidence that Class 0 objects have X-ray emission.
The deeply embedded sources with the strongest accretion activity are detected in the hard-band ($>$ 2keV) only. Due to extinction, their soft X-rays are not detected. To explain the decline in energy, one could consider that within a timescale of few Myrs the corona cools down via the accretion material, as seen in the pre-main-sequence Herbig AeBe stars \citep{mh08}. As a result, the older YSO (Class II/III) emit only in the soft-band.

\bigskip
\bigskip
\bigskip
\bigskip



\begin{figure}[ht!]
 \centering
 \includegraphics[angle=-90,width=0.5\textwidth]{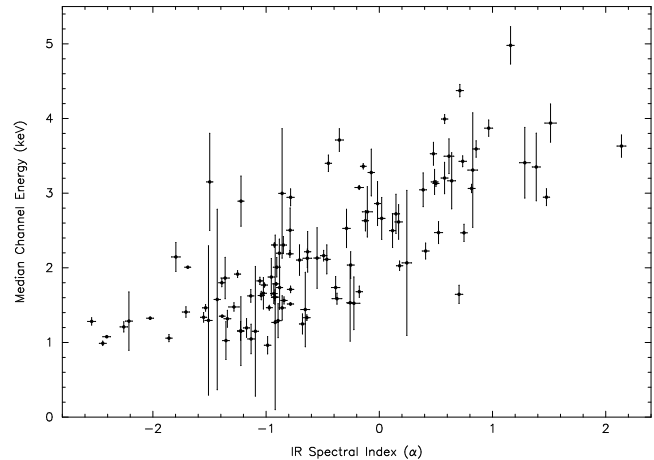}      
  \caption{X-ray energy of the YSO as a function of their infrared indices, showing an almost-linear correlation. This indicates that during the evolution of the YSOs, their X-ray energy becomes softer.}
  \label{fig1}
\end{figure}

\end{document}